\documentclass[12pt,a4paper]{article}

 \usepackage[dvips]{graphics}
 \usepackage{graphicx}
 \usepackage{afterpage}
 \usepackage{enumerate}
  \usepackage{longtable}
\begin{document}

 \title{\bf Depths of formation of magnetically sensitive absorption
lines }

 \author{\bf V.A. Sheminova}
 \date{}

 \maketitle
 \thanks{}
\begin{center}
{Main Astronomical Observatory, National Academy of Sciences of
Ukraine
\\ Zabolotnoho 27, 03689 Kyiv, Ukraine\\ E-mail: shem@mao.kiev.ua}
\end{center}

 \begin{abstract}
Characteristics of the depression contribution functions are studied for the Stokes
line profiles formed in a magnetic field. The form of the depression functions
depends mainly on the strength of splitting and the Zeeman component intensity, and
is of a complicated character with a distinctly pronounces asymmetry. The depths of
formation of magnetically sensitive lines are found by means of these contribution
functions. The calculations reveal that the steep section of the line profile is
formed higher than the profile center when a strong longitudinal magnetic field is
present. The Stokes profiles that describe the polarization characteristics are
formed only several kilometers higher than the Stokes profile that specifies the
general depression of the unpolarized and polarized radiation. The averaged depth
of formation of the whole line profile is practically independent of the magnetic
field strength. The depths of formation of 17 photospheric lines usually used in
magnetospectroscopic observations are calculated for the models of the quiet
photosphere, a flux tube, and the sunspot umbra.

\end{abstract}

\section{Introduction}
     \label{S-Introduction}

The determination of depths of formation of the observed Stokes parameter profiles
for magnetically sensitive Fraunhofer lines still remains a topical problem of the
spectral analysis which is used for obtaining information on the fine structure of
magnetic field in the solar photosphere. The problem of the depth of line formation
may be considered completely solved for the Fraunhofer lines which are formed in
the regions without a magnetic field, in the so-called quiet photosphere, while the
present state of the problem dealing with the depths of formation of the Stokes
line profiles still remains unsatisfactory.

In observational programs for studying the magnetic field structure in the
photospheric layers of the Sun, approximate estimates are usually used for
selecting lines with necessary characteristics. Thus, for example, to determine the
depth of formation of a Fraunhofer line, its equivalent width is estimated. It is
considered that the larger the equivalent width the higher in the atmosphere the
line is formed. The magnetic and temperature sensitivity of a line is determined by
the value of such atomic parameters of the line as the excitation potential,
ionization potential, and Land\'{e} factor. The smaller the sum of potentials the
more sensitive the line is to the temperature, and the greater the Land\'{e}
factor the more magnetically sensitive the line is. Though it is possible to get
approximate estimates from such reasonings, more reliable information on the depths
of formation of the Fraunhofer line profiles and the Stokes parameter profiles is
necessary for a detailed investigation of relationships between the physical
parameters of the atmosphere and the depth.

It is not long ago that we have begun the theoretical investigation of the Stokes
parameters of magnetically sensitive Fraunhofer lines. The calculation algorithm
for the Stokes profiles, for contribution and response functions was developed on
the basis of the theory of line formation in a magnetic field for the condition of
the local thermodynamic equilibrium; the theory was developed first by Unno and
improved by Rachkovskii, Beckers, and Landi Degl'Innocenti. The algorithm and
structure of the calculation program SPANSATM can be found in our paper
\cite{Sheminova90}. This program was produced on the basis of the SPANSAT program
\cite{Gadun88} intended for calculation and analysis of line profiles in stellar
atmospheres. The program system SPANSATM enables to carry out the complete analysis
of magnetically active photospheric lines in the solar spectrum. Our paper
\cite{Sheminova91} gives the results of investigations of the effect of physical
conditions in a magnetic medium and of atomic parameters of lines on the shape of
the Stokes profiles and on the magnetic broadening of the Fraunhofer lines. In the
present paper, the attention is concentrated mainly on the study of properties of
the depression contribution functions and on determination of the depths where the
Stokes profiles of photospheric lines are formed. The results of analysis of the
response functions of the Stokes parameters will be given in a next paper.

\section{Depression contribution functions
}

When the depths of line formation are calculated, the contribution functions which
describe the contribution of atmospheric layers to the observed value of relative
absorption in the spectral line are usually used. Besides the contribution
functions, the response functions are used in the interpretation of observations.
They serve for determination of the response of an observed quantity to a given
disturbance. These two functions supplement well each other and together they can
give a complete enough information about physical conditions in the medium where
spectral line profiles are formed. Depression contribution functions are the
functions which show what part of the radiation emerging at the surface (relative
to the continuum intensity $I_c$) is absorbed by each of the atmospheric layers in
a given region $\Delta \lambda$ of the spectral line. This function defines the
value of the contribution to the relative absorption $R( \Delta \lambda )=[I_c - I(
\Delta \lambda )] {/} { I_c }$. Here $I(\Delta \lambda)$ is the intensity of the
radiation emerging in the region $\Delta \lambda$ of the line. The depression
contribution function is equal to zero when the absorption coefficient in the line
is zero and the source function in the line $S_l$ is not equal to the intensity of
continuous radiation, i.e., $S_l$ should be greater or smaller than $I_c$. In other
words, the depression contribution function is significant if there are atoms which
absorb in the given line and the intensities of re-emitted and absorbed radiation
are different.

It is no mere chance that we have chosen the depression contribution functions as a
tool for investigating the regions where the line absorption processes are
localized and for determining the depths of formation of absorption lines. This
choice is based on the results of Gurtovenko and Sheminova \cite{Gurtovenko83} and
conclusions of Magain \cite{Magain}, where a formal solution of the radiation
transfer equation was obtained for a relative line depression in the integral form
and where the integrand was shown to be the only correct depression contribution
function fit for determining the depth of formation of an absorption line.
Gurtovenko and  Sarychev \cite{Gurtovenko88} demonstrate that a depression
contribution function proposed earlier which was derived using the
Uns\"{o}ld-Pecker procedure of weight functions is identical to the depression
contribution function from \cite{Magain}. At present, the depression contribution
function proposed first by Gurtovenko, Ratnikova, and de Jager \cite{Gurtovenko74}
and substantiated theoretically by Magain \cite{Magain} is widely used in
calculating the depths of formation of absorption lines and seems to be beyond any
doubt by now.

\section{Depression contribution functions \\ of the Stokes parameters
}

Very few studies have been dedicated to the contribution functions of the Stokes
parameters and to depths of their formation due to difficulties in calculations of
the Stokes profiles. The depths of formation of the Stokes profiles were studied
for the first time by Rachkovskii \cite{Rachkovskii69}, who compared the variations
with depth of the emergent radiation intensities in continuum and in lines and drew
the conclusion that the polarization parameters of radiation refer to higher layers
than the line intensity does. The contribution functions were already used by
Staude \cite{Staude72} to determine the depth of formation of magnetically
sensitive lines, but only the contribution to line emission was considered. These
studies had some drawbacks -- the depths found in them were not the depths of
absorption line formation, but characterized the regions of effective radiation in
a given line. In order to determine the depth of line formation in the presence of
a magnetic field, it is necessary to obtain the depression contribution functions
for the Stokes parameters and to calculate the depth of formation, using these
functions. The progress in studying the depths of formation of the Stokes
parameters began with paper \cite{Ballegooijen85} by van Ballegooijen, though the
depths of formation were not determined directly there. The method proposed in
\cite{Ballegooijen85} for solving the transfer equation for polarized radiation
enabled to write a single equation in a matrix form instead of four transfer
equations for the Stokes parameters; in that equation, the polarized radiation was
represented as the matrix

\[\mathbf{D}=\left(
  \begin{array}{cc}
    I+Q & U+iV \\
    U-iV & I-Q
  \end{array}
 \right),\]
where $I$, $Q$, $U$, $V$ are the Stokes parameters describing the intensity and
polarization characteristics of radiation. A formal solution of the transfer
equation was obtained in the integral form:

\[\mathbf{D}(0) = \int_{- \infty}^{\infty}  \mathbf{C}( \tau ) d \tau,\]
where the integrand $\mathbf{C}(\tau)$ is a matrix whose elements define the
contribution functions for the Stokes parameters. These contribution functions are
called the emission contribution functions of the Stokes parameters. They describe
the contribution of radiation of atmospheric layers to the polarized radiation
emerging in the line, and the integral of the contribution function along height
gives the value of the Stokes parameters. Thus, the author of \cite{Ballegooijen85}
succeeded in representing each of the Stokes parameters as an integral of an
ordinary function along height. This paper stimulated the deriving of the
depression contribution functions for the Stokes parameters of the Fraunhofer
lines. Grossman-Doerth et al. \cite{Grossman88}, using the method of
\cite{Ballegooijen85} and the procedure of \cite{Magain}, were the first to compose
the transfer equation for the relative depression of propagating polarized
radiation, this depression being represented as the matrix

\[\mathbf{\mathbf{R}}=0.5\left(
 \begin{array}{cc}
    R_I+R_Q & R_U+iR_V \\
    R_U-iR_V & R_I-R_Q
  \end{array} \right).
\]
The matrix elements in $\mathbf{R}$ contain the Stokes parameters of relative
depression: $R_I = 1 - I{/}I_c$; $R_Q = 1 - Q{/}I_c$; $R_U = 1 - U{/}I_c$; $R_V = 1
- V{/}I_c$. A formal solution of the transfer equation for the matrix $\mathbf{R}$
is the integral

\[\mathbf{R}(0) = \int_{- \infty}^{\infty}  \mathbf{C_R}( \tau ) d \tau,\]
where the integrand $\mathbf{C_R}(\tau)$ is also a matrix. The depression
contribution functions (similar to the emission ones) for each of the Stokes
parameters are defined in terms of the  $\mathbf{C_R}(\tau)$ matrix elements.
Mathematical expressions of the contribution functions for the Stokes parameters
can be found in \cite{Grossman88,Sheminova90}. We used the depression contribution
functions \cite{Grossman88} for the Stokes parameters in our study to determine the
depths of formation of the Stokes profiles of magnetically active absorption lines,
denoting them as $C_{RI}$, $C_{RQ}$, $C_{RU}$,  $C_{RV}$  in accordance with the
Stokes parameters.

\section{Characteristic features of the depression \\ contribution functions
for the Stokes parameters }

In a general case, a spectral line in a magnetic field splits into three groups of
components. The first group is characterized by absorption of the radiation
polarized linearly in the direction of the magnetic field (the $ \pi$ components).
The second group (the $\sigma_r$ components) is characterized by absorption of the
radiation right-hand circularly polarized in the plane perpendicular to the
direction of the field. The components of the third group (the $\sigma_b$
components) are left-hand circularly polarized in the same plane. The observed
relative line depression described by the Stokes parameter $R_I$ which represents
the absorption of both unpolarized and polarized light in a certain section of the
line profile contains information on all the components. The profile of $R_I$ is to
be considered as a superposition of profiles of all three component groups. In
consequence of this, the function of contribution to the relative depression
$C_{RI}$ may be of a complex character. As the contribution function $C_{RI}$
contains contributions of all components, its shape will depend on the spacing
between the groups of components and on their intensities. Since the $\pi$
components remain undisplaced in the magnetic field, the spacing between the
$\sigma$ groups which is determined according to the splitting rule is of
importance. The intensities of each group of components depend on the relative
selective absorption coefficient at the line center, on the inclination of magnetic
field, and on the value of anomalous dispersion. Thus, the contribution function
$C_{RI}$  is determined by the effective Land\'{e} factor $g_{\rm eff}$, magnetic
field strength, wavelength, number of absorbing atoms, Doppler width, damping
constant, magnetic field inclination, etc. The number of parameters is so great
that a detailed investigation of contribution functions is very laborious. It seems
reasonable to study only characteristic properties of the contribution functions.
With that end in view, we calculated the contribution functions for the Stokes
parameters of a line and separately for each group of components depending on the
profile section $\Delta\lambda$, the inclination angle $\gamma$, and the magnetic
field strength $H$. While calculating contribution functions for one group of
components, we assumed other functions to be equal to zero. The spectral line Fe I
643.085~nm was selected for calculations. As a model of the region where the line
absorption is formed, we took the empirical model of the magnetic flux tube
\cite{Walton87} which was obtained by comparing calculated profiles of the Stokes
parameters $I$ and an observed line profile in a plage area. The microturbulent
velocity was assumed to be 1~km/s, the damping constant was 1.5$\gamma_{vdW}$, and
the iron abundance was 7.64 dex.

\begin{figure}
   \centering
   \includegraphics[width=4.3 cm]{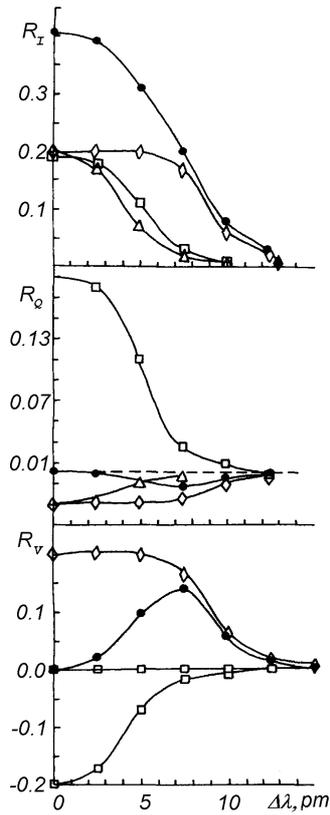}
   \hfill
\parbox[b]{10.7cm}{ \vspace{0.0cm}
   \caption[]{The Stokes profiles of the line Fe I
643.08~nm (points) and its $\pi$, $\sigma_b$, and $\sigma_r$ components (squares,
triangles, and diamonds, respectively) calculated for model \cite{Walton87}. }}
      \label{Fig1}
\end{figure}

\begin{figure}
   \centering
   \includegraphics[width=12. cm]{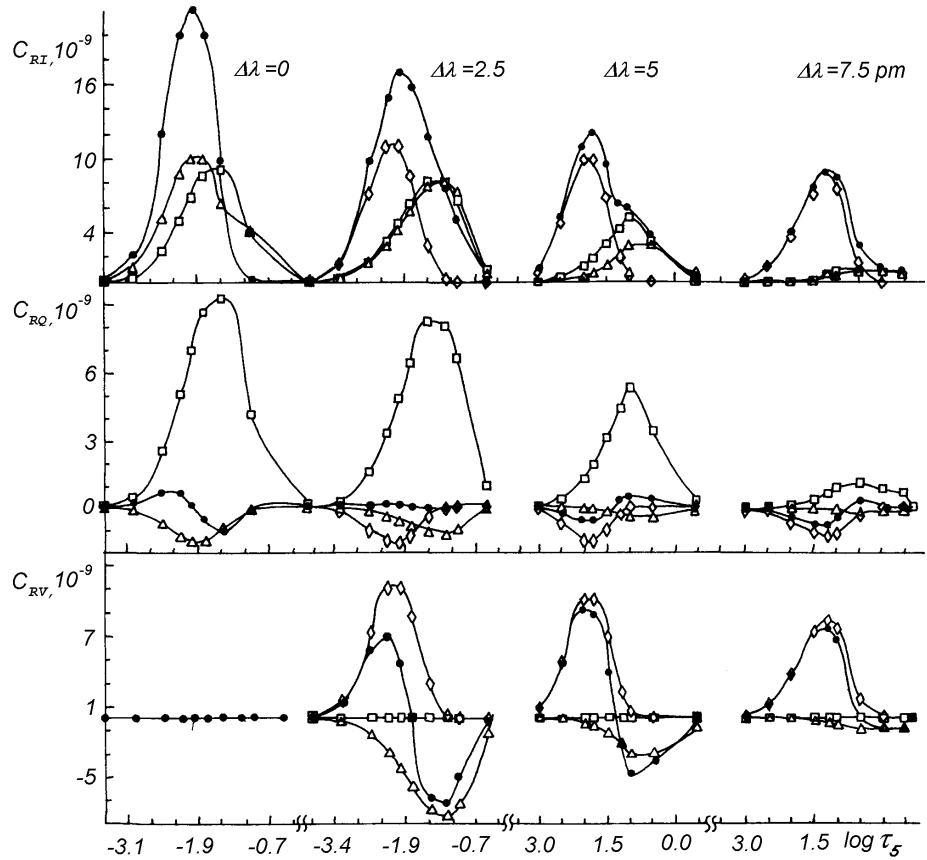}
    \caption[]{Depression contribution functions of the Stokes parameters for
different sections of line profiles (the notation is the same as in Fig.~1).
 }
      \label{Fig2}
\end{figure}

Analysis of the results obtained revealed a clear dependence of the contribution
function profiles for the Stokes parameters on the intensity of the Zeeman
components. Figures 1 and 2 show the profiles for the parameters $R_I$, $R_Q$,
$R_V$ and the corresponding contribution functions $C_{RI}$, $C_{RQ}$, and $C_{RV}$
for the line and for its groups of components at different profile sections
$\Delta\lambda$ which are formed in a magnetic flux tube with a magnetic field
strength of 0.1 T, an inclination of $30^{\circ}$ , and an azimuth of $0^{\circ}$.
It is evident that the contribution functions for the parameter $R_I$ are
asymmetric. A second maximum is outlined in those functions $C_{RI}$ which were
calculated for a steep part of the line profile, for $\Delta \lambda = 5$~pm, for
example. The contribution functions $C_{RI}$ have a shape close to a symmetric one
only for the center and for distant wings of the line. The contribution functions
of parameters $R_I$, $R_Q$, and $R_V$ can take on the negative values also, as they
characterize the difference in absorptions of polarized radiation for two
directions. The positive and negative maxima of the $C_{RQ}$ function change places
in going from the line center to the wing, which is evidence of a change in the
direction of light polarization along the line profile. The positive and negative
maxima of the contribution function $C_{RV}$ characterize the predominance of
contribution to the absorption of the right-hand and left-hand circularly polarized
light, respectively. Their contributions are equal at the line center, and so
$C_{RV}$  is equal to zero here. In going to the line wing, the value of the
$C_{RV}$ maxima varies in accordance with the contribution of the $\sigma$
components.

One can see in Figs 3 and 4 how the Stokes profiles $R_I$, $R_Q$, $R_V$ and their
contribution functions calculated for the profile sections $\Delta\lambda=0$, 5, 10
pm vary when the inclination increases from 0$^\circ$ to 90$^{\circ}$. The function
$C_{RI}$ is the most asymmetric for $\gamma=0^{\circ}$ and $\Delta \lambda=5$~pm.
The contribution of the undisplaced $\pi$ component grows with increasing $\gamma$.
This ``smooths'' the $C_{RI}$ contribution function, and its asymmetry diminishes.
The maximum values of the $C_{RQ}$ function become the greatest at
$\gamma=90^{\circ}$ and  $\Delta\lambda=5$~pm. The greatest positive and negative
maxima of the $C_{RV}$ function are observed at $\gamma=0^{\circ}$,
$\Delta\lambda=5$~pm and$\gamma=60^{\circ}$, $\Delta\lambda = 5$~pm, respectively.

\begin{figure}
   \centering
   \includegraphics[width=4.3 cm]{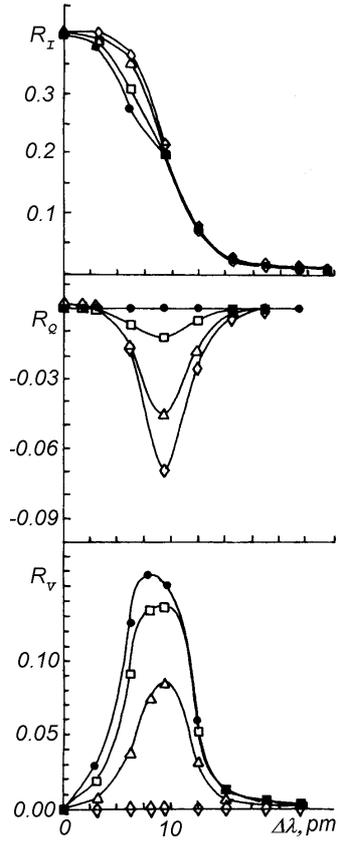}
   \hfill
\parbox[b]{10.5cm}{ \vspace{0.0cm}
   \caption[]{The Stokes profiles of the line Fe I
643.08~nm calculated for model \cite{Walton87} v. inclination: $ \gamma =0^{\circ}$ (points),
$30^{\circ}$(squares), $60^{\circ}$(triangles), $90^{\circ}$ (diamonds).
 } \label{Fig3}
 }

\end{figure}

\begin{figure}
   \centering
   \includegraphics[width=12. cm]{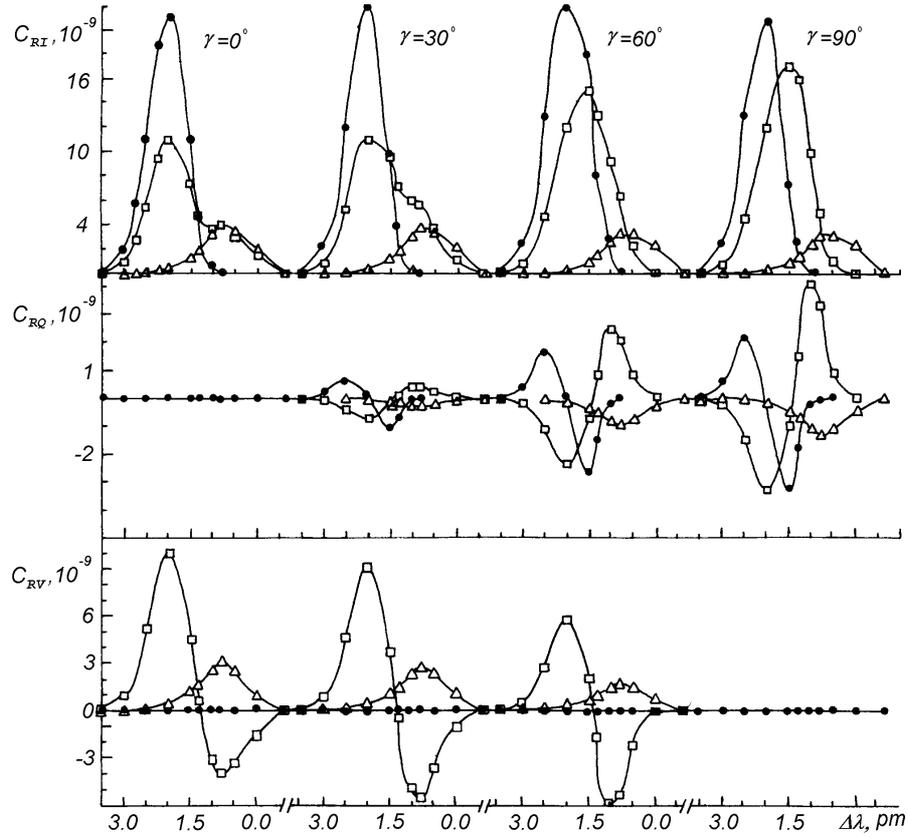}
    \caption[]{ Depression contribution functions of the Stokes
parameters for the  Fe I 643.08~nm line v. magnetic field inclination for $\Delta
\lambda =0$ (points), 5~pm (squares), 10~pm (triangles).
 }
      \label{Fig4}
\end{figure}

The effect of the magnetic field strength on the shape of profiles and contribution
functions for the sections $\Delta\lambda= 0$, 5, 10~pm is demonstrated in Figs 5
and 6. The profile shape for $R_I$ varies with increasing field strength. The
$C_{RI}$ functions for $\Delta\lambda$ = 0, 5~pm shift slightly to deeper layers of
the atmosphere. Their width increases, and the depression grows for $\Delta\lambda
=  10$~pm with growing $H$, and the values of maxima in the functions $C_{RI}$ grow
correspondingly. In the functions $C_{RV}$, the negative maximum diminishes with
increasing $H$ at the same distances from the line center, as the $\sigma$ group
moves away from the line center, and this means that its intensity becomes less at
the same $\Delta\lambda$. An increase of the field strength favours a ``clearing''
of the right-hand polarized absorption from the left-hand polarized one in the
right wing of the line profile.

It should be noted that the value of the Land\'{e} factor also affects strongly the
shape of the contribution functions. The larger  $g_{\rm eff}$ the greater is the
splitting of components, and the cases are possible when the components separate
entirely, i.e., a complete splitting occurs, and then the shape of the contribution
functions does not differ from a usual symmetric shape. For example, the complete
splitting is observed for the line Fe I  525.02~nm with $g_{\rm eff} =3$ which
originates in magnetic flux tubes with a longitudinal magnetic field of a strength
higher than 0.1~T.

So, variations of the contribution functions of magnetically active absorption
lines under the effect of a magnetic field correspond to variations of line
profiles. The contribution functions of the Stokes parameter $R_I$ which refer to
steep parts of wings of line profiles become asymmetric due to an increase of the
distance between splitted line components. A second maximum may appear for
moderately strong lines in which  $g_{\rm eff} \geq 2$ when the magnetic field is
longitudinal or is close to longitudinal and its strength is higher than 0.1 T. The
contribution functions for the polarization parameters $R_Q$,  $R_U$, and $R_V$
have a complex shape with a positive and a negative maxima.

\begin{figure}
   \centering
   \includegraphics[width=4.3 cm]{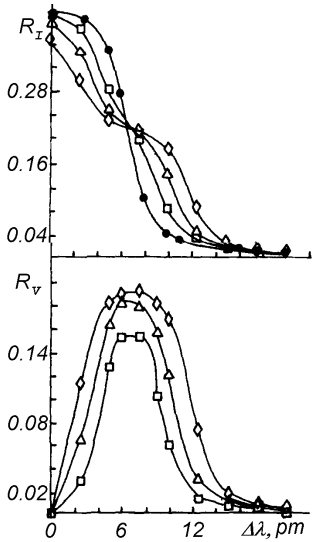}
   \hfill
\parbox[b]{8.5cm}{ \vspace{0.0cm}
   \caption[]{The Stokes profiles of the Fe I 643.08
nm  line  calculated for the strength $H=0.0$~T (points),  0.1~T (squares), 0.15~T
(triangles), 0.2~T (diamonds) of the longitudinal magnetic field in a magnetic flux
tube \cite{Walton87}.
 } \label{Fig5}
 }

\end{figure}
\begin{figure}
   \centering
   \includegraphics[width=9.5 cm]{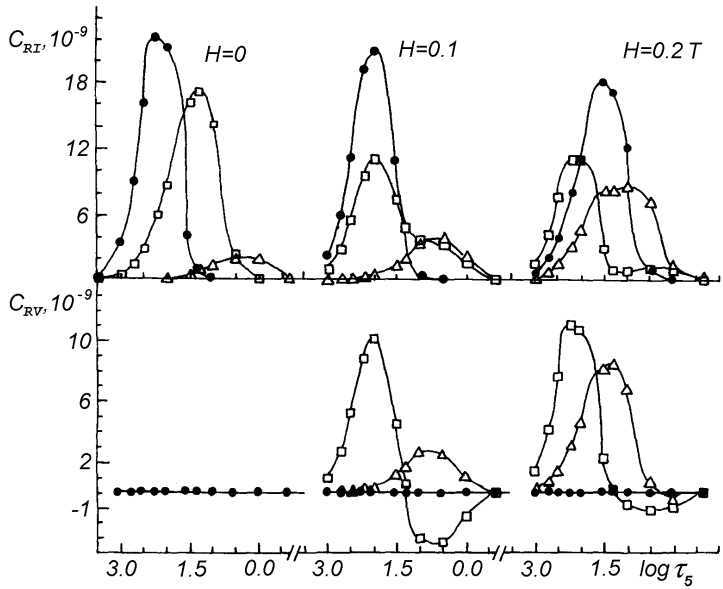}
   \hfill
\parbox[b]{6.cm}{ \vspace{0.7cm}
    \caption[]{Depression contribution functions of the Stokes
parameters for the  Fe I 643.08~nm line v. strength of the longitudinal magnetic
field for $\Delta \lambda =0$ (points), 5~pm (squares), 10~pm (triangles).
 }   \label{Fig6}
 }

\end{figure}

\section{Determining the depth of formation of the $R_I$ Stokes profile. }

The mean depth of the layer where the processes of selective absorption in a given
frequency interval of a spectral line go effectively is called the depth of
formation of the absorption line profile section, and it is calculated as a center
of gravity or as a weighted mean value:

\[h( \Delta \lambda )=\int_{- \infty }^{\infty}  h F(h, \Delta \lambda )dh /
\int_{- \infty }^{\infty}  F(h,  \Delta \lambda )dh.\] The depression contribution
function is used for the weight function $F$. The averaging operation is quite
justifiable here; this follows from \cite{Staude72}, for example, where it is shown
that, if a contribution function positive over the whole height interval is
normalized, then the new function

\[\varphi( h, \Delta \lambda )= F(h,  \Delta \lambda ){/} \int_
{- \infty }^{\infty}  F(h, \Delta \lambda )dh \] thus obtained may be considered as
the height distribution of probability density for the given process in the
atmosphere. Having calculated the center and variance of the distribution, one can
use this function for finding not only the depth but the width also of the layer
where a significant contribution occurs. The analysis of the contribution functions
for the Stokes parameters made by us has shown that the contribution function for
the parameter $R_I$ can be transformed to the probability density function. The
depression function $C_R$ is positive and it is easily normalized by dividing it by
the value of relative depression $R(\Delta\lambda)$. As a result, we obtain the
function

\[\varphi(h , \Delta \lambda ) = C_{RI} (h , \Delta \lambda )
{/} R ( \Delta \lambda ), \]
where

\[R ( \Delta \lambda )= \int_{- \infty }^{ \infty} C_{RI} ( h,  \Delta \lambda)dh ;\]
this function characterizes the probability of absorption of the unpolarized and
polarized radiation in a given line in a given spectral interval $ \Delta \lambda$.
The depth of formation $h_{RI}$ and the half-width of the effective layer $h_{RI}$
can be calculated now for the parameter $R_I$ by the following expressions:

\[h_{RI}( \Delta \lambda ) = \int_ {- \infty }^{\infty} h \varphi (h , \Delta \lambda
) dh,\]

\[\Delta  h_{RI}( \Delta \lambda ) = \int_ {- \infty }^{\infty} (h-h_{RI})^2 \varphi (h , \Delta \lambda
) dh.\]
Besides that, the so-called averaged depth of formation of the complete
line profile can be calculated as a weighted mean quantity:

\[h_W =  (1 {/} W) \int h_{RI} (   \Delta \lambda )  R_I ( \Delta \lambda )
d (  \Delta \lambda ),\]
where

\[W = \int   R_I ( \Delta \lambda ) d ( \Delta \lambda ).\]
The relative depression, i.e., the function $R_I( \Delta \lambda)$ is a weight
function here; $W$ is the total line depression, or the equivalent width.
Calculating the quantities $h_{RI}\pm \Delta h_{RI}$ and $h_W$ by the expressions
given here, we define completely the region of formation of the Stokes profile
$R_I$ and in that way solve the problem of the depth of formation of a magnetically
active absorption line in the presence of a magnetic field.

\section{Determining the depths of formation \\ of the polarization line characteristics
$R_Q$, $R_U$, $R_V$}

The formulae given above cannot be used in this case, since the contribution
functions for the Stokes parameters $R_Q$, $R_U$, and $R_V$ can take on both
positive and negative values in the height range considered here. In order to find
the mean depth of formation of a polarization parameter, it is more convenient to
resort to the center of gravity of the plane figure formed by the curve
corresponding to the contribution function and by abscissa. For example, let us
consider the contribution function $C_{RV}$ for the parameter $R_V$. The area of
the figure $S_{RV}$ is equal to the sum of two areas formed by the positive and
negative parts of the $C_{RV}$ curve which lie above and below the abscissa,
i.e.,$S_{RV} = S_1 + S_2$, where

\[S_1 = \left | \int_{-\infty }^{h_1}  C_{RV}(h, \Delta \lambda ) dh \right |,
S_2 = \int_{h_1}^{ \infty} C_{RV} ( h,\Delta \lambda ) dh,\] $h_1$ is the geometric
height at which the function $C_{RV}$ changes its sign. We shall use the expression

\[h_{ RV}= {\int \int} h dS_{RV} {/} S_{RV}\]
to determine the depth of formation of the $R_V$ profile. Substituting the
expression for $S_{RV}$ and transforming to a form convenient for calculations, we
get

\[h_{RV}= \int_{- \infty }^{\infty} h \left| C_{RV}
( \Delta \lambda ) \right | dh {/}  \int_{- \infty }^ {\infty} \left |
C_{RV}(\Delta \lambda ) \right | dh.\] The depths of formation for the parameters
$R_Q$ and $R_U$ can be found in the same way. Thus, to calculate the depths of
profile formation for polarization characteristics of magnetically active lines,
one can use the formula of weighted mean where the modulus of the depression
contribution function of the Stokes parameters is used for the weight function.

\section{Investigating depths of formation \\ of the Stokes profiles
}

Using three lines Fe I  525.02, 523.29, 643.08~nm as an example (Fig.~7), we
illustrated the variations in the depth of formation of the Stokes parameters with
varying magnetic field strength. We calculated the profiles and the depths of their
formation, varying the magnetic field strength $H$ from 0 to 0.2 T. Figure 8 shows
the effect of the field inclination on the depth of formation of the Stokes
profiles for the line Fe I 643.08~nm. When the inclination varies, the azimuth
remains equal to zero, and the field strength is 0.1 T. When calculating the depths
of formation of the Stokes profiles, we adopted the magnetic flux tube model of
\cite{Walton87} and used the geometric scale height $h$ and the logarithmic optical depth
scale log $\tau_5$, as these scales are mostly used in practice. The depths of
formation in the figures are given in the geometric scale height. The geometric
height zero level $h = 0$ corresponds to the optical depth $\tau_5 = 1$. Geometric
heights below the zero level take on negative values and above it they are
positive. Analysis of the results brings us to the following conclusions.

\begin{figure}
   \includegraphics[width=9.8 cm]{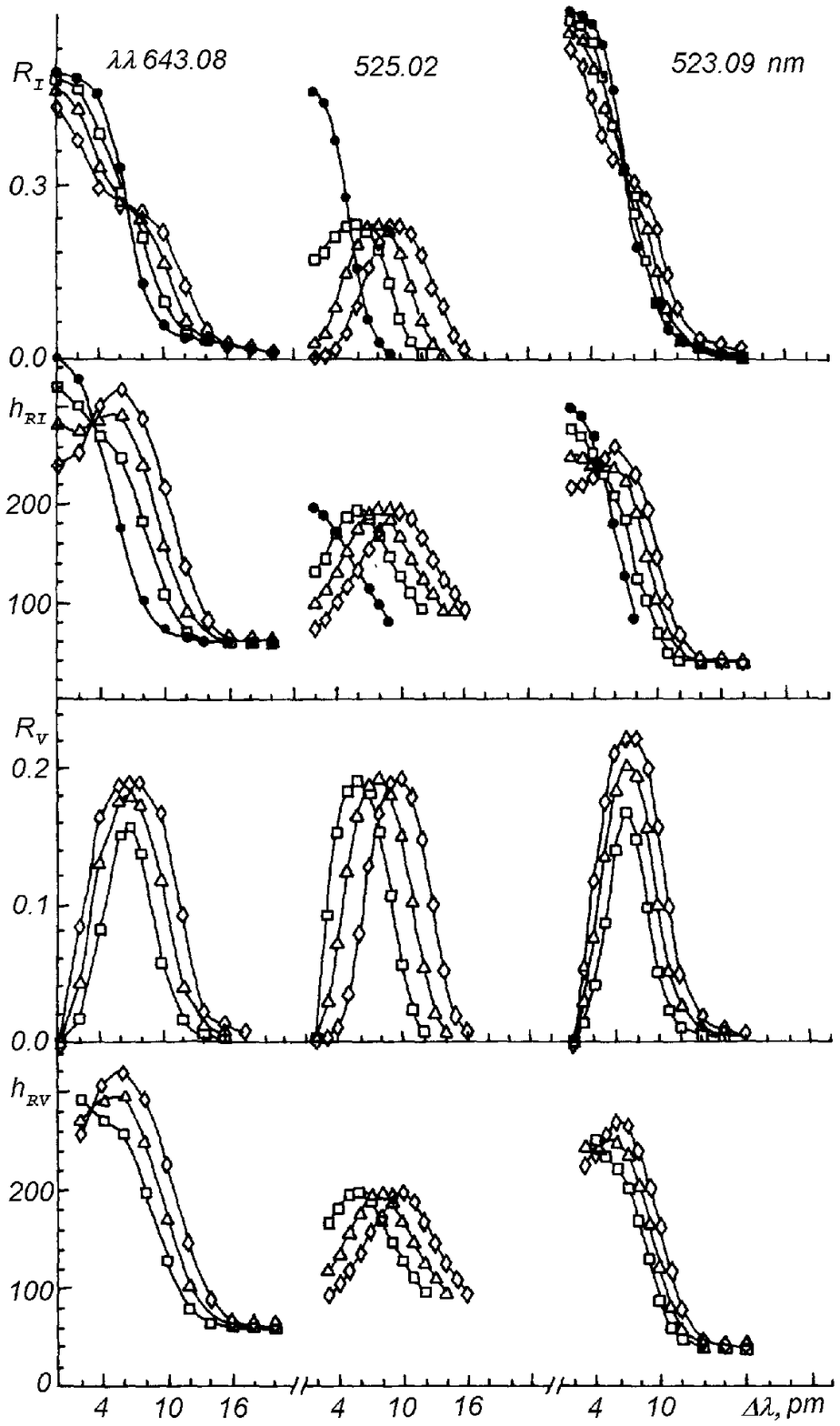}
    \includegraphics[width=5 cm]{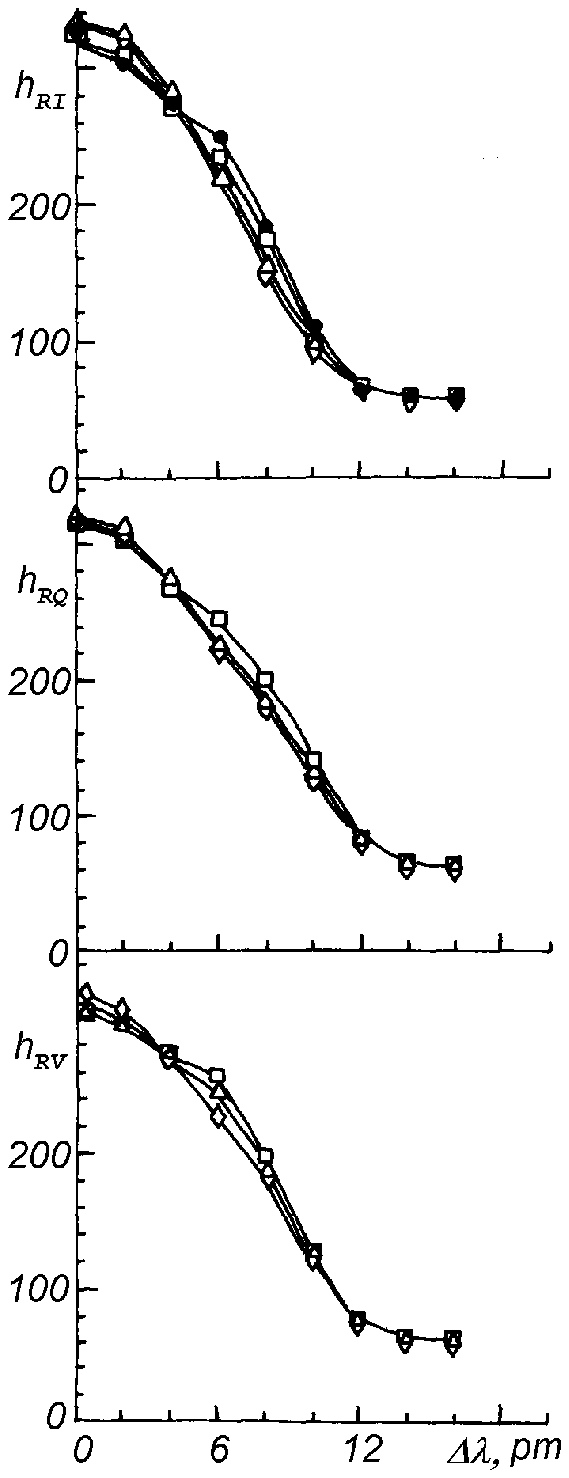}
   \hfill
\parbox[b]{9.cm}{ \vspace{0.7cm}
    \caption[]{The Stokes parameter profiles of three Fe~I
lines and depths of their formation $h_{RI}$ and $h_{RV}$ for magnetic field
strength $H=0$ (points), 0.1~T (squares), 0.15~T (triangles), 0.2 T (diamonds) of
the longitudinal magnetic field in a magnetic flux tube \cite{Walton87}.
 }
 }
   \hfill
\parbox[b]{5cm}{ \vspace{0.5cm}
   \caption[]{Depths of formation of the Stokes profiles
for the Fe I 643.08~nm line v. inclination. For the corresponding Stokes profiles
and notation, see Fig.~3.
 }}

\end{figure}

\section{On the depth of the layer of the $R_I$ profile \\
effective formation }

As is generally known, the Fraunhofer line profiles change their shape in the
presence of a magnetic field. It differs from the classic profile shape of a
Fraunhofer line (see Fig.~7). The depression in the central part of the profile
decreases with increasing field strength as compared to the profile calculated
without a magnetic field (a ``nonmagnetic'' profile), and in the line wings it
increases. The greater the Land\'{e} factor the stronger this effect is. The
profiles and the depth of formation of the depression change. It is evident from
Fig.~7 that, as the magnetic field strength  and the Land\'{e} factor are growing,
the central part of the profile is formed in more deep layers (the geometric
heights decrease), the middle part is formed in more high layers (the geometric
heights increase) as compared to the ``nonmagnetic'' profile, and far wings are
formed in the layers where the far wings of ``nonmagnetic'' profiles are formed.
The depth of line formation averaged over the entire profile depends on the manner
in which the equivalent width varies. The greater the magnetic broadening of the
line the higher in the atmosphere the entire line profile is formed on the average,
i.e., the magnetic field as if ``forces out'' slightly the line upwards. A
noticeable growth of the height may be expected for the lines with $g_{\rm eff}>2$
and $W> 7.0$~pm for $H> 0.1$~T. But calculations show that this growth is several
kilometers only. When magnetic fields are measured from the Fraunhofer lines, it
should be remembered that a steep section of the magnetic line profile can form
higher in the atmosphere than the profile center. Changes in the magnetic field
inclination from $0^{\circ}$ to $90^{\circ}$ affect little both the profile $R_I$
(Fig.~3) and the depths of its formation (Fig.~8). Appreciable changes are observed
only at the distances corresponding to the profile half-widths, where the geometric
height of formation decreases with increasing inclination.

\section{On the width of the layer of the $R_I$ profile \\
effective formation }

The width of the layer where the main fraction of polarized radiation is absorbed
in a certain part of line depends on the value of the line splitting in the
magnetic field, $ \Delta\lambda_H$. When the splitting grows, the layer width also
grows. The complete splitting having been reached, the layer width diminishes. The
width is different for different parts of the line. The layer width reaches its
maximum value, as a rule, for the middle part of the profile wings. The layer width
decreases when the magnetic field inclination varies from $0^{\circ}$ to
$90^{\circ}$.  Changes of the layer width that occur under the effect of a magnetic
field are as large as tens of kilometers. For example, for a region $\Delta\lambda
= 5$~pm in the line Fe I 643.08~nm which is formed in a magnetic flux tube with a
field strength of 0, 0.1, 0.15, 0.2, 0.3 T and $\gamma = 30^{\circ}$, $\varphi =
0^{\circ}$, the maximum half-width of the effective layer reaches 90, 123, 126,
119, 105~km, respectively. The maximum half-width for inclinations of $0^{\circ}$,
$30^{\circ}$, $60^{\circ}$, $90^{\circ}$ and $H = 0.1$ T is 123, 116, 100, 92~km,
respectively.

\begin{figure}
   \centering
   \includegraphics[width=11 cm]{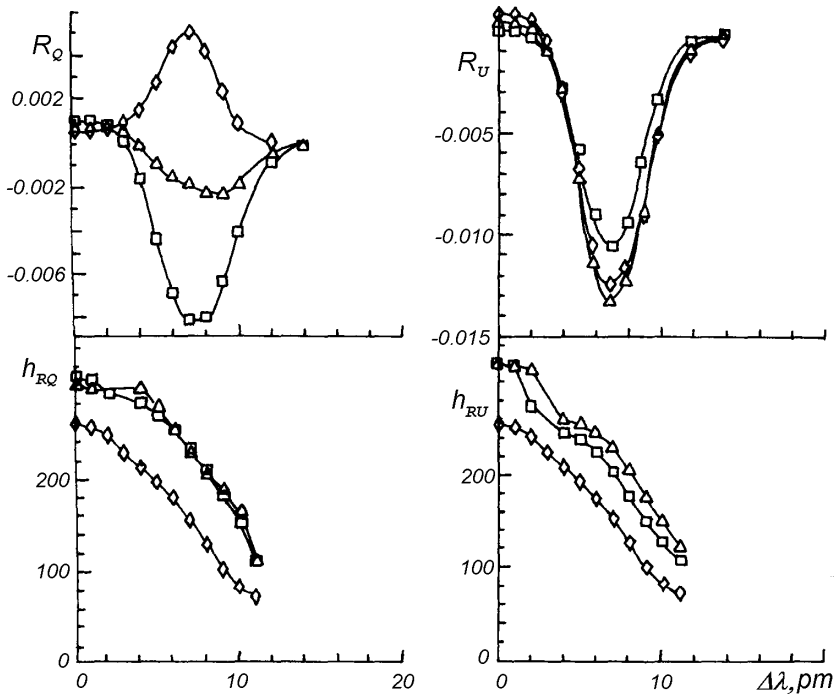}
   \hfill
\parbox[b]{4.cm}{ \vspace{0.7cm}
    \caption[]{Dependence of the Stokes profiles and depths of formation
on azimuth of magnetic field vector: $\varphi=15^{\circ}$  (sguares), $30^{\circ}$
(triangles), $45^{\circ}$ (diamonds) for flux tube model \cite{Walton87}.
 }  \label{Fig9}
 }

\end{figure}

\section{On the depths of formation of the polarization \\ line characteristics
$R_Q$, $R_U$, $R_V$  }

If the magnetic field is longitudinal, only one polarization parameter $R_V$
exists. The depth of its formation is found to be close to the depth of formation
of the $R_I$ profile (Fig.~7). The difference between them reaches several
kilometers, $R_V$ being formed slightly higher than $R_I$. Thus, the depths of
formation of the Stokes profiles $R_I$ and $R_V$ for a region $\Delta\lambda=5$~pm
of the lines Fe I 525.02, 523.29, 643.08~nm in a magnetic flux tube with $H =
0.1$~T, $\gamma= 0^{\circ}$,  $\varphi= 0^{\circ}$ are 187 and 188, 186 and 201,
262 and 267~km, respectively. If the magnetic field inclination
 $\gamma$ varies from  $0^{\circ}$ to  $90^{\circ}$
at  $\varphi= 0^{\circ}$, the parameter $R_Q$ appears in addition to $R_V$, and its
profile is formed higher that $R_I$ (Fig.~8). For example, for $\Delta\lambda = 5$
pm of the line 643.08~nm in a magnetic flux tube with $H = 0.1$ T,
 $\gamma= 30^{\circ}$ , $\varphi= 0^{\circ}$ , the
$R_I$, $R_Q$, and $R_V$ profiles are formed at depths of 253, 258, 256~km,
respectively. If we change the angle $\varphi$ , the parameter $R_U$ will appear.
Figure 9 shows how the depths of formation of $R_Q$ and $R_U$ change in these
cases. It follows from the results of calculations that the polarization profiles
$R_Q$, $R_U$, and $R_V$ are formed higher than the general relative depression of
the line, $ R_I$, but the difference is insignificant on the average. Therefore, it
may be assumed that all the Stokes profiles are formed in the same layers.

\section{The depths of formation of lines  frequently \\ used in magnetographic
observations   }

For calculating the depths of line formation, first of all it is necessary to
define physical conditions in the region of formation or to specify a model solar
atmosphere. It is well known that even the simplest horizontally homogeneous model
atmosphere is defined by a large number of parameters which vary with height. The
accuracy of calculations depends therefore completely on the accuracy of simulation
of the conditions in the atmospheric regions where the processes of absorption line
formation go. The calculated depth of line formation refers directly to the
specific model used in the calculations. Variations of physical conditions in the
atmosphere cause variations of such line parameters as the central intensity,
equivalent width, profile shape, and the depth of formation as well. Such a notion
as the depth of line formation appears to become more indeterminate and becomes
meaningless in inhomogeneous model atmospheres. In this case, it is better to use
the concept of the effective region of formation.


%
{\footnotesize
 \begin{table}[t] \centering
 \parbox[b]{14cm}{
\caption{Parameters of spectral lines used in calculations. \label{T:1}}
\vspace{0.3cm}}
 \footnotesize
\begin{tabular}{clrrrrr}
 \hline
 Line number (no.) & $\lambda $, nm & $EP$, eV & $W_{obs}$, pm  &

$R$ & $\log gf$ &$g_{\rm eff}$ \\
 \hline
 1 & 480.815 Fe I   &   3.25        &    2.74       &  0.336         &   -2.69        &  1.33 \\
 2 & 523.294 Fe I   &   2.93        &    21.00      &  0.803         &   -1.62        &  1.30 \\
 3 & 523.462 Fe II  &   3.22        &    8.85       &  0.722         &   -2.31        &  0.93 \\
 4 & 524.705 Fe I   &   0.09        &    6.04       &  0.716         &   -5.03        &  2.00 \\
 5 & 524.757 Cr I   &   0.96        &    7.83       &  0.749         &   -1.62        &  2.50 \\
 6 & 525.021 Fe I    &   0.12        &    6.48       &  0.710         &   -4.89        &  3.00 \\
 7 & 609.666 Fe I   &   3.98        &    3.91       &  0.358         &   -1.85        &  1.50 \\
 8 & 611.165 ~V I   &   1.04        &    1.09       &  0.084         &   -0.68        &  1.33 \\
 9 & 612.622 Ti I    &   1.07        &    2.10       &  0.222         &   -1.37        &  1.25 \\
 10& 612.897 Ni I    &   1.68        &    2.36       &  0.262         &   -3.27        &  1.50 \\
 11& 615.162 Fe I    &   2.18        &    4.85       &  0.507         &   -3.37        &  1.83 \\
 12& 617.334 Fe I    &   2.22        &    6.93       &  0.622         &   -2.90        &  2.50 \\
 13& 630.252 Fe I   &   3.69        &    9.08       &  0.650         &   -1.14        &  2.50 \\
 14& 630.346 Fe I   &   4.32        &    0.47       &  0.045         &   -2.67        &  1.50 \\
 15& 636.946 Fe II  &   2.89        &    2.07       &  0.172         &   -4.28        &  2.10 \\
 16& 643.085 Fe I   &   2.18        &    11.17      &  0.725         &   -2.08        &  1.25 \\
 17& 673.316 Fe I    &   4.64        &    2.74       &  0.239         &   -1.51        &  2.50 \\
    \hline
 \end{tabular}
 \end{table}}
 \noindent

To demonstrate the aforesaid, we chose three substantially different models. They
are the models of the quiet photosphere (HOLMU \cite{Holweger74}), a plage area
(WAL2 \cite{Walton87}), and a spot (OBST \cite{Obridko88}). Table 1 gives the
parameters of the lines used. The chemical element abundances were taken according
to \cite{Gurtovenko89}. The magnetic field is longitudinal with a strength of 0.15
T and the azimuth  $\varphi= 0^{\circ}$. The microturbulent velocity is 1~km/s
(macroturbulence was ignored), the damping constant is $1.5 \gamma_{vdW}$. Tables
2--5 give the results of calculations for the depths of formation of selected lines
for each of the models. Table 2 contains the calculation data for the quiet
photosphere without regard to a magnetic field. The values of the following
parameters are given for each line: equivalent width $W$ averaged over the entire
line profile, depth of formation $h_W$, optical depth $\log \tau_{5W}$, selected
section on the profile $\Delta\lambda$, relative line depression $R$, parameters
$R_I$ and $R_V$, depth of formation and half-width of the effective layer of the
depression $\log \tau_{5RI}\pm \Delta$, $h_{RI}\pm \Delta$, $h_{RV}\pm \Delta$. In
order to demonstrate variations in the depth of formation over a profile from one
model to another, we selected three sections on the profile calculated for the
HOLMU model without a magnetic field: at the line center, at a distance
corresponding to the line half-width, and in a wing, where the depression was 1\%.
Further, the depths of formation were calculated in other models for the same
sections $\Delta\lambda$. An additional point on the profile was calculated in that
part of the wing where it became wider. The data given in the tables demonstrate
clearly the reaction of lines to the conditions of their formation. As an example,
we may point at the lines whose equivalent width more than doubled upon the
temperature rise. These are the lines V I 611.16~nm, Ti I 612.62~nm, Ni I
612.89~nm, FeI  524.75, 525.02, 615.16, 617.33, 630.25~nm. The lines Fe II 523.46
and 636.94~nm react weakly to the temperature rise, but they practically disappear
when the temperature drops, in a spot, for example. Other conclusions may be also
drawn from the data given in the tables, and everyone who is interested in and
needs this has such an opportunity.

 %


{\footnotesize
 \begin{longtable}{cccccccc}
 \multicolumn{8}{c}{Table 2. Depths of line formation. Quiet photosphere model
  HOLMU.  $H=0$~T. }
 \vspace{0.3cm}\\
 \hline
 no.& $W$, nm & $h_W$, km & $\log \tau_{5\,W}$  &
 $\Delta\lambda $, pm &$~~R~~$& $\log \tau_5 \pm \Delta~$ & $~h\pm\Delta$, km \\
 \hline\endfirsthead
   \multicolumn{8}{c}{\hspace{10.5cm}Table 2 (continued)}\\
    \vspace{0.3cm}\\
   \hline
 no.& $W$, nm & $h_W$, km & $\log \tau_{5\,W}$  &
 $\Delta\lambda $, pm & $~~R~~$ & $\log \tau_5 \pm \Delta$ & $~h\pm\Delta$, km \\
 \hline\endhead
   \hline
    \multicolumn{8}{c}{\hspace{12cm}}
    \endfoot
\hline\endlastfoot
 1 &  2.68  &   166   &  -1.11  &   0 &  0.48 &   -1.25~~   0.75 &   187~~ 116    \\
   &        &         &         &   3 &  0.20 &   -0.98~~   0.72 &   145~~ 111    \\
   &        &         &         &   6 &  0.12 &   -0.62~~   0.59 &   91~~ 89     \\
 2 &  9.66  &   296   &  -1.96  &   0 &  0.80 &   -2.85~~   0.69 &   431~~ 104    \\
   &        &         &         &   5 &  0.52 &   -1.22~~   0.71 &   183~~ 110    \\
   &        &         &         &  20 &  0.01 &   -0.63~~   0.53 &   92~~ 81     \\
 3 &  8.86  &   262   &  -1.74  &   0 &  0.78 &   -2.59~~   0.79 &   392~~ 120    \\
   &        &         &         &   6 &  0.28 &   -0.62~~   0.59 &   90~~ 89     \\
   &        &         &         &  17 &  0.01 &   -0.34~~   0.48 &   50~~ 68     \\
 4 &  6.28  &   315   &  -2.08  &   0 &  0.77 &   -2.54~~   0.85 &   384~~ 129    \\
   &        &         &         &   4 &  0.39 &   -1.55~~   0.85 &   233~~ 130    \\
   &        &         &         &   8 &  0.01 &   -0.97~~   0.65 &   143~~ 101    \\
 5 &  7.59  &  300    &-1.98    &   0 &  0.78 &   -2.57~~   0.74 &   390~~ 111    \\
   &        &         &         &   5 &  0.33 &   -1.20~~   0.74 &   179~~ 115    \\
   &        &         &         &   11&  0.01 &   -0.76~~   0.56 &   111~~ 86     \\
 6 &  6.74  &  328    &-2.17    &   0 &  0.79 &   -2.69~~   0.84 &   408~~ 127    \\
   &        &         &         &   4 &  0.46 &   -1.60~~   0.85 &   241~~ 131    \\
   &        &         &         &   8 &  0.01 &   -0.96~~   0.65 &   142~~ 101    \\
 7 &  3.78  &  177    &-1.18    &   0 &  0.48 &   -1.38~~   0.73 &   208~~ 112    \\
   &        &         &         &   4 &  0.22 &   -1.00~~   0.69 &   148~~ 107    \\
   &        &         &         &   9 &  0.01 &   -0.56~~   0.52 &   82~~ 78     \\
 8 &  1.08  &  189    &-1.26    &   0 &  0.17 &   -1.32~~   0.78 &   197~~ 120    \\
   &        &         &         &   3 &  0.09 &   -1.23~~   0.76 &   184~~ 118    \\
   &        &         &         &   6 &  0.01 &   -1.05~~   0.72 &   156~~ 112    \\
 9 &  2.04  &  197   -&1.31     &   0 &  0.30 &   -1.40~~   0.79 &   210~~ 122    \\
   &        &         &         &   3 &  0.17 &   -1.27~~   0.77 &   191~~ 119    \\
   &        &         &         &   7 &  0.0l &   -0.97~~   0.70 &   145~~ 108    \\
 10&  2.30  &  201    &-1.34    &   0 &  0.35 &   -1.44~~   0.81 &   217~~ 125    \\
   &        &         &         &   3 &  0.19 &   -1.28~~   0.79 &   192~~ 122    \\
   &        &         &         &   7 &  0.01 &   -0.93~~   0.69 &   138~~ 106    \\
 11&  4.83  &  234    &-1.55    &   0 &  0.59 &   -1.81~~   0.79 &   274~~ 121    \\
   &        &         &         &   4 &  0.30 &   -1.30~~   0.77 &   195~~ 119    \\
   &        &         &         &   8 &  0.01 &   -0.83~~   0.62 &   122~~ 96     \\
 12&  6.97  &  279    &-1.85    &   0 &  0.68 &   -2,33~~   0.75 &   353~~ 113    \\
   &        &         &         &   5 &  0.35 &   -1.29~~   0.76 &   193~~ 117    \\
   &        &         &         &   11&  0.01 &   -0.74~~   0.55 &   108~~ 85     \\
 13&  9.30  &  267    &-1.77    &   0 &  0.69 &   -2.51~~   0.68 &   380~~ 102    \\
   &        &         &         &   6 &  0.37 &   -1.10~~   0.67 &   163~~ 104    \\
   &        &         &         &   30&  0.01 &   -0.60~~   0.51 &   88~~ 78     \\
 14&  0.45  &  138    &-0.93    &   0 &  0.07 &   -0.97~~   0.70 &   144~~ 108    \\
   &        &         &         &   3 &  0.04 &   -0.90~~   0.69 &   133~~ 106    \\
   &        &         &         &   5 &  0.01 &   -0.79~~   0.65 &   116~~ 100    \\
 15&  1.99  &   132   & -0.90   &   0 &  0.28 &   -1.00~~   0.76 &   148~~ 117    \\
   &        &         &         &   3 &  0.17 &   -0.85~~   0.73 &   125~~ 112    \\
   &        &         &         &   7 &  0.01 &   -0.55~~   0.62 &   81~~ 93     \\
 16&  11.83 &   358   & -2.37   &   0 &  0.75 &   -3.46~~   0.70 &   522~~ 108    \\
   &        &         &         &   7 &  0.42 &   -1.29~~   0.72 &   193~~ 111    \\
   &        &         &         &   30&  0.01 &   -0.76~~   0.55 &   111~~ 85     \\
 17&  2.72  &  154    &-1.04    &   0 &  0.33 &   -1.18~~   0.70 &   176~~ 109    \\
   &        &         &         &   4 &  0.16 &   -0.93~~   0.67 &   138~~ 104    \\
   &        &         &         &   9 &  0.01 &   -0.54~~   0.52 &   78~~ 77     \\
 \end{longtable}}

%

{\footnotesize
 \begin{longtable}{cccccccccc}
 \multicolumn{10}{c}{ Table 3. Depths of line formation. Quiet photosphere model HOLMU.
 $H=0.15$~T.
 \label{T:3} } \vspace{0.3cm}\\
 \hline
 no.& $W,$\,nm & $h_W,$\,km & $\log \tau_{5W}$  &
$\Delta\lambda$,\,pm& $R_I$ &$\log \tau_{5RI}\pm\Delta~$&
$~h_{RI}\pm\Delta$,\,km&$R_V~$& $h_{RV}\pm\Delta$,\,km \\
 \hline\endfirsthead
   \multicolumn{10}{c}{\hspace{12.cm}Table 3 (continued)} \vspace{0.3cm}\\
   \hline
 no.& $W,$\,nm & $h_W,$ \,km & $\log \tau_{5W}$  &
$\Delta\lambda$,\,pm & $R_I$ &$\log \tau_{5RI}\pm\Delta~$&
$h_{RI}\pm\Delta$,\,km&$R_V~$& $h_{RV}\pm\Delta$,\,km \\ \hline\endhead
   \hline
    \multicolumn{10}{c}{\hspace{12cm}}
    \endfoot
\hline\endlastfoot
 1 &  2.69  &166& -1.11  & 0 &  0.31 &   -1.08~~ 0.73&  161~~  113  &   0.00  &             \\
   &        &   &        & 3 &  0.24 &   -1.18~~ 0.75&  178~~  116  &   0.21  &  187~~  115    \\
   &        &   &        & 8 &  0.01 &   -0.63~~ 0.59&   92~~   89   &   0.01  &  96~~   92     \\
 2 &  9.76  &298& -1.97  & 0 &  0.77 &   -2.40~~ 0.70&  365~~  105  &   0.00  &     ~~         \\
   &        &   &        & 5 &  0.44 &   -2.17~~ 0.90&  329~~  136  &   0.33  &  336~~  137    \\
   &        &   &        & 20&  0.01 &   -0.63~~ 0.53&   92~~   81   &   0.003 &  92~~   81     \\
 3 &  8.88  &262& -1.74  & 0 &  0.76 &   -2.33~~ 0.79&  354~~  119  &   0.00  &     ~~         \\
   &        &   &        & 6 &  0.34 &   -1.21~~ 0.75&  181~~  116  &   0.26  &  196~~  116    \\
   &        &   &        & 17&  0.01 &   -0.34~~ 0.48&   50~~   68   &   0.003 &  50~~   68     \\
 4 &  6.46  &312& -2.06  & 0 &  0.46 &   -1.62~~ 0.85&  244~~  132  &   0.00  &     ~~         \\
   &        &   &        & 4 &  0.39 &   -2.49~~ 0.87&  377~~  132  &   0.38  &  382~~  129    \\
   &        &   &        & 8 &  0.19 &   -1.54~~ 0.85&  232~~  131  &   0.19  &  233~~  131    \\
   &        &   &        & 12& 0.004 &   -0.95~~ 0.63&  140~~   99  &   0.002 &  141~~  100    \\
 5 &  7.62  &300& -1.98  & 0 &  0.38 &   -1.24~~ 0.75&  187~~  116  &   0.00  &     ~~         \\
   &        &   &        & 5 &  0.40 &   -2.53~~ 0.77&  384~~  116  &   0.38  &  387~~  115    \\
   &        &   &        & 11&  0.10 &   -1.04~~ 0.71&  155~~  110  &   0.09  &  158~~  111    \\
   &        &   &        & 14&  0.01 &   -0.75~~ 0.56&  111~~   87  &   0.01  &  111~~   87    \\
 6 &  6.78  &330& -2.18  & 0 &  0.09 &   -1.23~~ 0.78&  184~~  121  &   0.00  &     ~~         \\
   &        &   &        & 4 &  0.38 &   -2.40~~ 0.86&  364~~  130  &   0.38  &  367~~  128    \\
   &        &   &        & 8 &  0.37 &   -2.27~~ 0.86&  344~~  131  &   0.37  &  345~~  130    \\
   &        &   &        & 12&  0.02 &   -1.09~~ 0.73&  162~~  113  &   0.02  &  164~~  114    \\
 7 &  3.97  &174& -1.17  & 0 &  0.26 &   -1.06~~ 0.71&  158~~  109  &   0.00  &     ~~         \\
   &        &   &        & 4 &  0.24 &   -1.29~~ 0.73&  193~~  113  &   0.22  &  203~~  111    \\
   &        &   &        & 9 &  0.07 &   -0.92~~ 0.68&  137~~  105  &   0.07  &  140~~  106    \\
   &        &   &        & 12&  0.01 &   -0.60~~ 0.54&   87~~   82  &   0.01  &   89~~   84     \\
 8 &  1.08  &189& -1.26  & 0 &  0.07 &  -1.21~~ 4.76 &  181~~  118  &   0.00  &     ~~         \\
   &        &   &        & 3 &  0.09 &  -1.30~~ 0.78 &  195~~  120  &   0.08  &  199~~  120    \\
   &        &   &        & 6 &  0.05 &  -1.25~~ 0.77 &  187~~  119  &   0.05  &  188~~  119    \\
   &        &   &        & 9 &  0.01 &  -1.08~~ 0.73 &  161~~  113  &   0.01  &  162~~  113    \\
 9 &  2.08  &196& -1.31  & 0 &  0.15 &  -1.27~~ 0.77 &  190~~  119  &   0.00  &     ~~         \\
   &        &   &        & 3 &  0.15 &  -1.36~~ 0.79 &  204~~  121  &   0.13  &  212~~  121   \\
   &        &   &        & 7 &  0.06 &  -1.24~~ 0.77 &  185~~  118  &   0.06  &  187~~  119    \\
   &        &   &        & 10&  0.01 &  -1.03~~ 0.72 &  154~~  111  &   0.01  &  157~~  112    \\
 10&   2.33 &200& -1.33  & 0 &  0.12 &  -1.22~~ 0.77  & 182~~  120   &  0.00   &    ~~          \\
   &        &   &        & 3 &  0.17 &  -1.40~~ 0.80 &  210~~  124   &  0.15   & 214~~  124     \\
   &        &   &        & 7 &  0.09 &  -1.28~~ 0.79 &  191~~  122   &  0.09   & 193~~  121     \\
   &        &   &        & 10&  0.01 &  -1.05~~ 0.73 &  156~~  113   &  0.01   & 158~~  114     \\
 11&   4.92 &233& -1.54  & 0 &  0.19 &  -1.19~~ 0.75 &  178~~  117   &  0.00   &    ~~          \\
   &        &   &        & 4 &  0.29 &  -1.75~~ 0.80 &  265~~  122   &  0.28   & 270~~  120     \\
   &        &   &        & 8 &  0.21 &  -1.44~~ 0.78 &  217~~  121   &  0.21   & 219~~  121     \\
   &        &   &        & 12&  0.01 &  -0.87~~ 0.65 &  128~~  100   &  0.01   & 131~~  102     \\
 12&   6.99 &281&  -1.86 & 0 & 0.11  &  -1.00~~ 0.70  & 148~~  108    & 0.00    &   ~~           \\
   &        &   &        & 5 &  0.33 &   -2.16~~ 0.77 & 327~~  117   &  0.33   & 331~~  115     \\
   &        &   &        & 11&  0.23 &   -1.45~~ 0.77 & 220~~  119   & 0.23    & 221~~  119     \\
   &        &   &        & 16&  0.01 &   -0.75~~ 0.56 & 111~~   87   & 0.01    & 111~~   88     \\
 13&   9.20 &270& -1.79  & 0 & 0.23  &   -0.88~~ 0.62  &129~~   96    & 0.00    &   ~~           \\
   &        &   &        & 6 &  0.36 &  -2.39~~ 0.77 &  362~~  116   &  0.33   & 364~~  115     \\
   &        &   &        & 25&  0.01 &  -0.60~~ 0.51 &   87~~   78   &  0.005  &  88~~   78     \\
 14&   0.44 &137& -0.92  & 0 & 0.02  &  -0.85~~ 0.67  & 125~~  103    & 0.00    &   ~~           \\
   &        &   &        & 3 &  0.03 &  -0.94~~ 0.70 &  140~~  108   &  0.03   & 145~~  108     \\
   &        &   &        & 5 &  0.03 &  -0.96~~ 0.70 &  142~~  108   &  0.03   & 144~~  108     \\
   &        &   &        & 9 &  0.01 &  -0.83~~  0.67 & 123~~  103   &  0.01   & 124~~  103     \\
 15&   1.99 &132& -0.90  & 0 &  0.03 &  -0.63~~  0.66 &  92~~   99   &  0.00   &    ~~          \\
   &        &   &        & 3 &  0.08 &  -0.85~~  0.73 & 125~~  112   &  0.08   & 127~~  112     \\
   &        &   &        & 7 &  0.13 &  -0.97~~  0.75 & 144~~  116   &  0.13   & 145~~  116     \\
   &        &   &        & 12&  0.01 &  -0.59~~  0.64 &  86~~   96   &  0.01   &  87~~   97     \\
 16&  12.10 &360& -2.38  & 0 &  0.71 &  -2.83~~  0.72 & 428~~  108   &  0.00   &    ~~          \\
   &        &   &        & 7 &  0.40 &  -2.68~~  0.96 & 405~~  145   &  0.31   & 409~~  146    \\
   &        &   &        & 30&  0.01 &  -0.76~~  0.55 & 111~~   85   &  0.002  & 112~~   85    \\
 17&  2.75  &155& -1.04  & 0 & 0.02  &  -0.59~~  0.55  & 86~~    83   &   0.00  &   ~~           \\
   &        &   &        & 4 &  0.08 &  -0.93~~  0.67 & 137~~   104  &   0.08  & 141~~  104   \\
   &        &   &        & 9 &  0.16 &  -1.15~~  0.70 & 172~~   109  &   0.16  & 173~~  109   \\
   &        &   &        & 16&  0.01 &  -0.55~~  0.53 &  80~~    79  &   0.01  &  81~~   79   \\
\end{longtable}}



 %
%
{\footnotesize
 \begin{longtable}{cccccccccc}
 \multicolumn{10}{c}{Table 4. Depths of line formation. Magnetic flux tube model
 of Walton \cite{Walton87}.  $H=0.15$~T.
 \label{T:2} } \vspace{0.3cm}\\
 \hline
 no.& $W$,\,nm & $h_W$,\,km & $\log \tau_{5W}$  &
 $\Delta\lambda$,\,pm & $R_I$ &$\log \tau_{5RI}\pm\Delta$& $h_{RI}\pm\Delta$,\,km&$R_V$&
 $h_{RV}\pm\Delta$,\,km \\
 \hline\endfirsthead
    \multicolumn{10}{c}{\hspace{12.cm}Table 4 (continued)} \vspace{0.3cm}\\
   \hline
 no.& $W$,\,nm & $h_W$,\,km & $\log \tau_{5W}$  &
 $\Delta\lambda$,\,pm & $R_I$ &$\log \tau_{5RI}\pm\Delta$& $h_{RI}\pm\Delta$,\,km&$R_V$&
 $h_{RV}\pm\Delta$,\,km \\
 \hline\endhead
   \hline
    \multicolumn{10}{c}{\hspace{12cm}}
    \endfoot
\hline\endlastfoot

  1 &  1.42 & 102& -0.71 &  0 &  0.16 &   -0.69~~   0.67  &   99~~    100    &0.00 &               \\
    &       &    &       &  3 &  0.12 &   -0.77~~   0.68  &  111~~   101    &0.11  &  121~~   100     \\
    &       &    &       &  8 &  0.01 &   -0.33~~   0.60  &   47~~    86    &0.01  &   52~~    89      \\
   2&   5.71& 210& -1.42 &  0 &  0.47 &   -1.71~~   0.53 &   252~~    90    &0.00  &     ~~           \\
    &       &    &       &  5 &  0.27 &   -1.53~~   0.69 &   227~~   111    &0.20  &  234~~   111    \\
    &       &    &       &  17&  0.01 &   -0.27~~    0.53&    38~~    75    &0.003 &   39~~    75     \\
   3&   5.94& 320& -1.98 &  0 &  0.44 &   -2.77~~    0.63&   461~~    48    & 0.00 &     ~~           \\
    &       &    &       &  6 &  0.26 &   -1.57~~    0.90&   240~~   157    & 0.19 &   255~~  158    \\
    &       &    &       & 17 &  0.01 &   -0.23~~    0.57&    33~~     82   &0.002 &    34~~  83       \\
   4&   2.60& 166& -1.15 &  0 &  0.14 &   -0.93~~    0.67&   134~~   102   & 0.00 &               \\
    &       &    &       &  4 &  0.19 &   -1.31~~   0.63 &   190~~    99   & 0.18 &    193~~     98  \\
    &       &    &       &  8 &  0.06 &   -0.89~~    0.68&   128~~   102  & 0.06 &     129~~     102 \\
    &       &    &       & 10 &  0.01 &   -0.69~~    0.67&    98~~     99  &0.01  &    101~~     100  \\
   5&   3.98& 187& -1.28 &  0 &  0.16 &   -0.79~~    0.66&   113~~    99  &0.00  &               \\
    &       &    &       &  5 &  0.23 &   -1.59~~   0.57 &   235~~     94  &0.23  &    237~~      93   \\
    &       &    &       &  11&  0.04 &   -0.62~~    0.65&    89~~     97  &0.04  &     91~~   97       \\
    &       &    &       &  12&  0.01 &   -0.39~~    0.58&    55~~     83  &0.01  &     56~~   84       \\
   6&  2.92 & 176& -1.21 &  0 &  0.03 &   -0.71~~    0.68&   102~~   100  &0.00  &        ~~        \\
    &       &    &       &  4 &  0.18 &   -1.28~~   0.63 &   185~~     99  &0.18  &    187~~      98   \\
    &       &    &       &  8 &  0.17 &   -1.22~~   0.64 &   176~~     99  &0.17  &    177~~      99   \\
    &       &    &       &  12&  0.01 &   -0.66~~    0.67&    94~~     98  &0.01  &     96~~   99       \\
 7  &  1.92 & 125& -0.87 &  0 &  0.12 &   -0.79~~    0.65&   113~~    98  &0.00  &        ~~        \\
    &       &    &       &  4 &  0.12 &   -0.98~~    0.64&   141~~    98  &0.10  &   151~~   94   \\
    &       &    &       &  9 &  0.04 &   -0.65~~    0.65&    94~~     97  &0.04  &   97~~   97       \\
    &       &    &       &  12&  0.01 &   -0.40~~    0.60&    56~~     86  &0.01  &   60~~   88       \\
 8  &  0.32 & 118& -0.82 &  0 &  0.02 &   -0.78~~    0.67&   112~~    101 & 0.00 &      ~~           \\
    &       &    &       &  3 &  0.03 &   -0.84~~    0.67&   121~~    101 & 0.02 &   125~~       101 \\
    &       &    &       &  6 &  0.02 &   -0.81~~    0.67&   117~~    101 & 0.02 &   118~~      101 \\
 9  &  0.66 & 122& -0.85 &  0 &  0.05 &   -0.82~~    0.67&   118~~    101 & 0.00 &      ~~          \\
    &       &    &       &  3 &  0.05 &   -0.88~~    0.67&   126~~    101 & 0.04 &   132~~      100 \\
    &       &    &       &  7 &  0.02 &   -0.80~~    0.67&   115~~    101 & 0.02 &   117~~      101 \\
 10 &  0.81 & 129& -0.90 &  0 &  0.04 &   -0.82~~    0.68&   117~~    101 & 0.00 &      ~~          \\
    &       &    &       &  3 &  0.06 &   -0.93~~    0.67&   134~~    102 & 0.05 &   138~~      101 \\
    &       &    &       &  7 &  0.03 &   -0.86~~    0.67&   124~~    102 & 0.03 &   124~~      102 \\
    &       &    &       &  10&  0.01 &   -0.72~~    0.67&   103~~    100 & 0.01 &   105~~      100 \\
 11 &  2.01 & 149& -1.03 &  0 &  0.07 &   -0.79~~    0.67&   114~~    100 & 0.00 &               \\
    &       &    &       &  4 &  0.13 &   -1.15~~   0.63 &   166~~      98 & 0.12 &   169~~      96  \\
    &       &    &       &  8 &  0.08 &   -0.97~~    0.65&   140~~      99 & 0.08 &   141~~      99  \\
    &       &    &       &  12&  0.004&   -0.53~~    0.63&    74~~      91 & 0.004&    78~~   93      \\
 12 &  3.30 & 183& -1.26 &  0 &  0.04 &   -0.64~~    0.65&    92~~      96 & 0.00 &      ~~          \\
    &       &    &       &  5 &  0.17 &   -1.41~~   0.58 &   206~~      94 & 0.17 &   209~~      93  \\
    &       &    &       &  11&  0.11 &   -1.09~~   0.63 &   157~~      97 & 0.12 &   158~~      96  \\
    &       &    &       &  14&  0.01 &   -0.51~~    0.62&    73~~      90 & 0.01 &    75~~   91      \\
 13 &  4.95 & 203& -1.38 &  0 &  0.11 &   -0.66~~    0.60&    93~~      89 & 0.00 &      ~~          \\
    &       &    &       &  6 &  0.21 &   -1.78~~   0.59 &   265~~      99 & 0.19 &   267~~      98  \\
    &       &    &       &  20&  0.01 &   -0.31~~    0.53&    44~~      75 & 0.01 &    45~~   75      \\
 14 &  0.19 &  94& -0.66 &  0 &  0.01 &   -0.58~~    0.66&    83~~      97 & 0.00 &      ~~          \\
    &       &    &       &  3 &  0.01 &   -0.68~~    0.67&    97~~      99 & 0.01 &   101~~     100 \\
    &       &    &       &  5 &  0.02 &   -0.69~~    0.67&    98~~     100 & 0.01 &   101~~      100 \\
    &       &    &       &  7 &  0.01 &   -0.64~~    0.67&    92~~      99 & 0.01 &    93~~   99      \\
 15 &  1.49 & 160& -1.08 &  0 &  0.02 &   -0.74~~    0.81&   109~~      128& 0.00 &      ~~          \\
    &       &    &       &  3 &  0.06 &   -1.02~~   0.82 &   151~~      133& 0.06 &   153~~      133 \\
    &       &    &       &  7 &  0.09 &   -1.18~~   0.82 &   175~~      134& 0.09 &   176~~      134 \\
    &       &    &       &  12&  0.01 &   -0.68~~    0.81&   100~~      125& 0.01 &   101~~      126 \\
 16 &  6.11 & 252& -1.68 &  0 &  0.39 &   -1.90~~   0.49 &   284~~      87 & 0.00 &      ~~          \\
    &       &    &       &  7 &  0.22 &   -1.82~~   0.64 &   273~~      107& 0.18 &   278~~      106 \\
    &       &    &       &  20&  0.01 &   -0.42~~    0.55&    59~~       79& 0.003&    61~~   79      \\
 17 &  1.35 & 117& -0.82 &  0 &  0.01 &   -0.36~~    0.58&    52~~       83&  0.00&      ~~          \\
    &       &    &       &  4 &  0.04 &   -0.71~~    0.65&    02~~     96&  0.04&      107~~      96 \\
    &       &    &       &  9 &  0.08 &   -0.93~~    0.64&    33~~     97&  0.07&      134~~      96 \\
    &       &    &       &  14&  0.01 &   -0.47~~    0.62&    67~~      89&  0.01&      69~~   90     \\
\end{longtable}}

 %
{\footnotesize
 \begin{longtable}{cccccccccc}
 \multicolumn{10}{c}{Table 5. Depths of line formation. Solar sunspot umbra model
of Obridko and Staude \cite{Obridko88}. $H=0.15$~T. } \vspace{0.3cm}\\
 \hline
 no.& $W$,\,nm & $h_W$,\,km&$\log \tau_{5W}$&
 $\Delta\lambda$,\,pm &$R_I$&$\log \tau_{5RI}\pm\Delta$&$h_{RI}\pm\Delta$,\,km&$R_V$&
 $h_{RV}\pm\Delta$,\,km \\
 \hline\endfirsthead
    \multicolumn{10}{c}{\hspace{12.cm}Table 5 (continued)} \vspace{0.3cm}\\
   \hline
 no.& $W$,\,nm& $h_W$,\,km &$\log \tau_{5W}$&
 $\Delta\lambda$,\,pm &$R_I$&$\log \tau_{5RI}\pm\Delta$& $h_{RI}\pm\Delta$,\,km&$R_V$&
 $h_{RV}\pm\Delta$,\,km \\
 \hline\endhead
   \hline
    \multicolumn{10}{c}{\hspace{12cm}}
    \endfoot
\hline\endlastfoot

 1 &    3.30&   33 & -0.54  &  0    & 0.32   &  -0.62~~   0.67  &    39~~  40   &0.00  &            \\
   &        &      &        &  3    & 0.25   &  -0.71~~    0.72 &    44~~  44  &0.16  &    56~~  43    \\
   &        &      &        &  8    & 0.04   &  -0.04~~    0.44 &     2~~  29  &0.02  &     4~~    30   \\
   &        &      &        &  14   & 0.01   &   0.02~~     0.43 &   -2~~  28  &0.003 &    -2~~   29   \\
 2 &   16.12&   64 & -1.08  &    0  &  0.73  &  -2.11~~    0.73&    125~~  40  &0.00  &      ~~        \\
   &        &      &        &   5   & 0.52   &  -1.63~~   0.98 &     98~~  56  &0.21  &    99~~   58   \\
   &        &      &        &   70  & 0.01   &  -0.07~~    0.51 &     3~~  33  &0.001 &    15~~   55   \\
 3 &   0.33 &  -7  &  0.08  &    0  &  0.04  &   0.07~~     0.48&    -6~~  32  &0.00  &      ~~        \\
   &        &      &        &   5   & 0.01   &   0.18~~     0.43 &  -13~~  29 & 0.01 &    -11~~  29   \\
 4 &   17.15&   123&  -2.11 &    0  &  0.81  &  -2.91~~    0.83&    168~~  46 & 0.00 &       ~~       \\
   &        &      &        &   4   & 0.50   &  -3.88~~    1.11 &   224~~  68 & 0.37 &    224~~  68   \\
   &        &      &        &   8   & 0.50   &  -2.24~~    1.18 &   131~~  67 & 0.29 &    134~~  68   \\
   &        &      &        &   60  & 0.01   &  -0.35~~    0.60 &    22~~  37 & 0.001&     22~~   37   \\
 5 &   33.27&   93 & -1.59  &    0  &  0.79  &  -2.61~~    0.74&    152~~  39 & 0.00 &       ~~       \\
   &        &      &        &   4   & 0.48   &  -4.24~~    0.80 &   243~~  50 & 0.39 &    243~~  51   \\
   &        &      &        &   8   & 0.55   &  -3.24~~    1.17 &   185~~  65 & 0.32 &    186~~  65   \\
   &        &      &        &   81  & 0.03   &  -0.33~~    0.55 &    20~~  35 & 0.004&     22~~   34   \\
 6 &   18.60&   118&  -2.03 &    0  &  0.56  &  -1.39~~   0.75&      85~~   42 & 0.00 &      ~~       \\
   &        &      &        &   5   & 0.45   &  -4.19~~    0.81 &   242~~   55 & 0.40 &   242~~  55   \\
   &        &      &        &   11  & 0.39   &  -1.55~~    0.93 &    93~~   53  &0.25  &   99~~   54   \\
   &        &      &        &   70  & 0.01   &  -0.34~~    0.59 &    21~~   37  &0.001 &   22~~   37   \\
 7 &   2.54 &  35  & -0.56  &    0  &  0.14  &  -0.53~~    0.63&     33~~   39 & 0.00 &      ~~       \\
   &        &      &        &   4   & 0.14   &  -0.72~~    0.68 &    44~~   41  &0.10  &   54~~   40   \\
   &        &      &        &   9   & 0.05   &  -0.33~~    0.57 &    21~~   36  &0.04  &   27~~   36   \\
   &        &      &        &   16  & 0.01   &  -0.04~~    0.44 &     2~~    29  &0.004 &   3~~    29   \\
 8 &   9.54 &  108 & -1.84  &    0  &  0.60  &  -2.02~~    0.83&    120~~    45 & 0.00 &     ~~        \\
   &        &      &        &   3   & 0.49   &  -2.33~~    1.14 &   136~~   63 & 0.25 &   138~~   65  \\
   &        &      &        &   6   & 0.39   &  -2.17~~    1.00 &   127~~    55&  0.29&   130~~   55  \\
   &        &      &        &   30  & 0.01   &  -0.40~~    0.57 &    25~~    35 & 0.002&   25~~   35   \\
 9 &   14.85&   116&  -1.98 &    0  &  0.75  &  -2.87~~    0.74&    165~~   39&  0.00&       ~~      \\
   &        &      &        &    3  &  0.64  &  -2.70~~    1.18&    155~~   64&  0.17&    155~~   68  \\
   &        &      &        &    7  &  0.46  &  -2.28~~    1.10&    133~~   61&  0.27&    135~~   62  \\
   &        &      &        &    50 &  0.01  &  -0.43~~    0.57&     27~~   35 & 0.001&    28~~   35   \\
 10&   3.29 &  64  & -1.05  &     0 &  0.18  &  -0.84~~    0.79&     52~~   47 & 0.00 &      ~~       \\
   &        &      &        &     3 &  0.21  &  -1.23~~   0.87&      75~~   50 & 0.18 &    83~~   48   \\
   &        &      &        &     7 &  0.13  &  -0.99~~    0.82&     61~~   48 & 0.12 &    64~~   47   \\
   &        &      &        &    14 &  0.01  &  -0.24~~    0.53&     15~~   34 & 0.003&    15~~   34   \\
 11&   7.15 &  80  & -1.32  &     0 &  0.31  &  -0.96~~    0.70&     59~~   41 & 0.00 &      ~~       \\
   &        &      &        &     4 &  0.34  &  -1.80~~   0.93&     107~~  53 & 0.26 &    110~~  53   \\
   &        &      &        &     8 &  0.26  &  -1.42~~   0.82&      86~~   47 & 0.22 &    90~~   46   \\
   &        &      &        &    26 &  0.01  &  -0.20~~    0.51&     12~~   33 & 0.003&    13~~   33   \\
 12&   10.53&   86 & -1.44  &     0 &  0.25  &  -0.70~~    0.56&     43~~   34 & 0.00 &      ~~       \\
   &        &      &        &     5 &  0.38  &  -2.12~~    1.01&    124~~   57 & 0.29 &   126~~  57   \\
   &        &      &        &    11 &  0.27  &  -1.36~~    0.79&     83~~   45 & 0.22 &    87~~   45   \\
   &        &      &        &    43 &  0.01  &  -0.20~~    0.50&     12~~   33 & 0.003&    13~~   32   \\
 13&   9.61 &   60 & -0.99  &     0 &  0.21  &  -0.55~~    0.46&     35~~   29 & 0.00 &      ~~       \\
   &        &      &        &     6 &  0.33  &  -1.63~~    0.86&     98~~   50 & 0.24 &   100~~  50  \\
   &        &      &        &    47 &  0.01  &  -0.07~~    0.46&      4~~    30 & 0.003&    5~~  30   \\
 14&   0.15 &  23  & -0.37  &     0 &  0.01  &  -0.25~~    0.58&     15~~   37 & 0.00 &      ~~       \\
   &        &      &        &     3 &  0.01  &  -0.40~~    0.66&     25~~   41 & 0.10 &    32~~  42   \\
   &        &      &        &     6 &  0.01  &  -0.39~~    0.65&     24~~   41 & 0.01 &    28~~  41   \\
 15&   0.004&   -4 &   0.04 &     0 & 0.0001 &   0.17~~     0.43 &  -12~~   29 & 0.00 &      ~~       \\
   &        &      &        &     3 & 0.0003 &   0.05~~     0.53 &   -5~~   35  &0.0003&   -4~~  36   \\
   &        &      &        &     5 & 0.0006 &   0.01~~     0.56 &   -2~~   37  &0.0000&   -1~~   37   \\
 16&   23.30&   93 & -1.57  &     0 &   0.73 &  -2.97~~   0.77&     172~~  44 & 0.00 &       ~~     \\
   &        &      &        &     7 &   0.52 &  -2.45~~   1.21&     143~~  68 & 0.23 &    144~~   69  \\
   &        &      &        &    80 &  0.02  &  -0.24~~    0.51&     15~~   33 & 0.001&    16~~   33   \\
 17&   0.99 &  26  & -0.42  &     0 &   0.02 &  -0.98~~   0.45&       6~~    29 & 0.00 &     ~~       \\
   &        &      &        &     4 &   0.03 &  -0.33~~   0.58&      21~~   37 & 0.02 &    27~~   37   \\
   &        &      &        &     9 &   0.05 &  -0.57~~   0.66&      35~~   40 & 0.05 &    38~~   40   \\
   &        &      &        &    16 &  0.01  &  -0.07~~    0.45&      4~~    29 & 0.01 &    5~~    29   \\
\end{longtable}}

\section{Conclusion}

The depression contribution functions for the Stokes parameters of magnetically
active absorption lines are of a complex nature in the conditions of the solar
atmosphere in the regions where a strong magnetic field is present. The depression
functions may be considered as a result of combining the contribution functions of
groups of the components into which a line splits in a magnetic field. The function
profiles depend to a large extent on the value of splitting $\Delta\lambda_H$ and
on the intensity of each of the components. The depression functions may be used to
determine the depth and width of the effective layer where the Stokes profiles are
formed.

The profiles of the line polarization parameters are formed several kilometers
higher than the profile of the parameter which describes the general depression of
polarized and unpolarized radiation. The Stokes profiles may be considered to form
in the same layer due to a small difference between the depths of their formation.
Magnetically active lines with large values of the Land\'{e} factor which form in
the presence of a strong longitudinal magnetic field have a distinctive feature --
the steep section of the line profile is formed higher than the center of the line
profile.

The averaged depth of line formation depends on the amount for magnetic broadening.
When the latter increases, the whole region of line formation shifts slightly into
the layers lying higher. This effect being insignificant, the averaged depth of
line formation in a magnetic field does not change.

It is not possible to draw up tables of calculated depths of formation of
magnetically active lines on account of a strong dependence of the value of
depression and the depth of line formation on physical conditions in the medium.
The depths of formation should be calculated every time the conditions in the
medium change.



\newpage

\end{document}